\documentclass[preprint,times]{elsarticle}
\pdfoutput=1

\usepackage{amsmath}  
\usepackage{amssymb}
\usepackage{subfigure}
\usepackage{ecrc}
\usepackage{rotating}
\usepackage{xcolor}
\usepackage{latexsym}
\usepackage{bm}
\usepackage{floatrow}
\newfloatcommand{capbtabbox}{table}[][0.45\textwidth]

\volume{}
\jid{}
\firstpage{1}
\journalname{}
\jnltitlelogo{}
\runauth{}

\newcommand{\be}{\begin{equation}}
\newcommand{\ee}{\end{equation}}
\newcommand{\ba}{\begin{eqnarray}}
\newcommand{\ea}{\end{eqnarray}}

\begin{document}
\begin{frontmatter}

\title{\textcolor{black}{Void growth and coalescence in irradiated copper\\ under deformation}}

\author[]{P.O. Barrioz\corref{cor1}}
\author[]{J. Hure}
\author[]{B. Tanguy}
\cortext[cor1]{Corresponding author}
\address[]{CEA Saclay, Universit\'e Paris-Saclay, DEN, Service d'\'Etudes des Mat\'eriaux Irradi\'es, 91191 Gif-sur-Yvette, France}

\begin{abstract}
  A decrease of fracture toughness of irradiated materials is usually observed, as reported \textcolor{black}{for austenitic stainless steels in Light Water Reactors (LWRs) or copper alloys for fusion applications}. \textcolor{black}{For a wide range of applications (\textit{e.g.} structural steels irradiated at low homologous temperature), void growth and coalescence fracture mechanism has been shown to be still predominant. As a consequence, a comprehensive study of the effects of irradiation-induced hardening mechanisms on void growth and coalescence in irradiated materials is required.} The effects of irradiation on ductile fracture mechanisms - void growth to coalescence - are assessed in this study based on model experiments. Pure copper thin tensile samples have been irradiated with protons up to 0.01~dpa. Micron-scale holes drilled through the thickness of these samples subjected to uniaxial loading conditions allow a detailed description of void growth and coalescence. \textcolor{black}{In this study, experimental data show that physical mechanisms of micron-scale void growth and coalescence are similar between the unirradiated and irradiated copper. However, an acceleration of void growth is observed in the later case, resulting in earlier coalescence, which is consistent with the decrease of fracture toughness reported in irradiated materials. These results are qualitatively reproduced with numerical simulations accounting for irradiation macroscopic hardening and decrease of strain-hardening capability.} 
\end{abstract}


\end{frontmatter}

\section{Introduction}

\begin{figure}[H]
\centering
\includegraphics[height = 5cm]{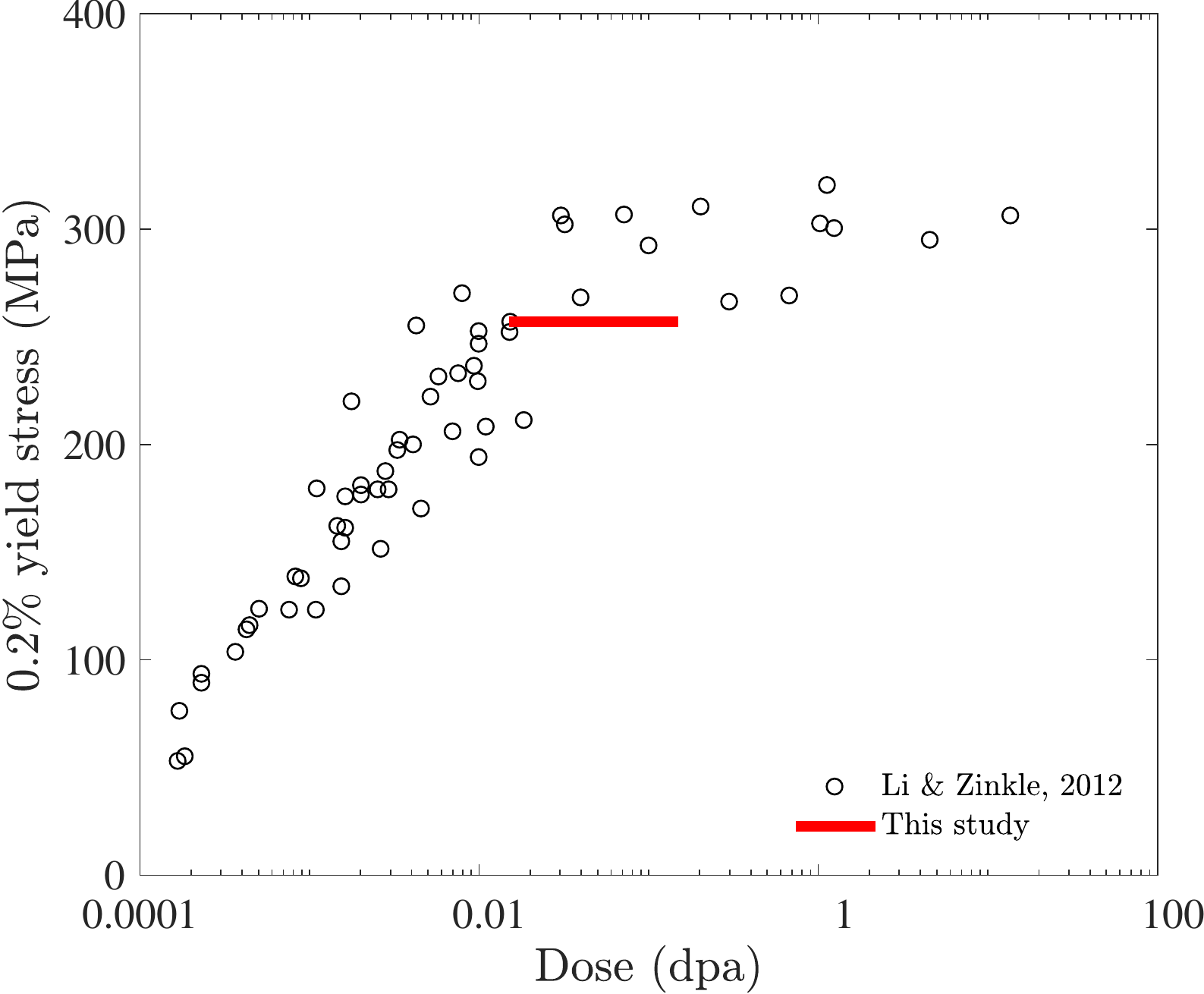}
\caption{\textcolor{black}{Evolution of conventional 0.2\% yield stress of pure copper with dose (data taken from \cite{reviewcu} and references therein). The red line corresponds to this study considering the range of dose in the depth of the irradiated layer (see Section~2.2).}}
\label{fig0}
\end{figure}

Structural materials used for fission reactor cores \cite{zinklechallenge} (or selected for ITER fusion reactor \cite{iter}) are subjected to high energy neutron irradiation and high irradiation dose, leading to significant evolutions of mechanical properties related to the creation of irradiation defects in the microstructure. 300 series austenitic Stainless Steels (SS) are used for Light Water Reactors (LWR) core internals, and are also foreseen for first wall/blanket and divertor of ITER fusion reactor. For the latter, copper alloys are also considered. Fracture toughness of these materials and its evolution with irradiation are required for design purposes, but also for ageing management, as experimental studies have shown a strong decrease of toughness with irradiation. Reviews of the degradation of austenitic stainless steels toughness with irradiation under LWR conditions can be found in \cite{epri,chopra1,fukuyareview}. A decrease of toughness (as measured through initiation energy release rate $J_{Ic}$) up to a factor ten is observed after a few dpa. Fracture surfaces of unirradiated SS exhibit transgranular dimples, indicating void nucleation \cite{argon75,beevers}, growth \cite{mcclintock,ricetracey} and coalescence mechanisms \cite{koplik}. Classically, voids nucleate by cracking or decohesion of inclusions or second-phase particles, then grow due to the plastic flow of the matrix material around them until strong interactions between adjacents voids appear, which correspond to the coalescence phase. Details about these mechanisms can be found elsewhere \cite{pineaureview}. \textcolor{black}{It should be noted that void growth and coalescence of concern here is due to plastic flow under mechanical loading post to irradiation, which is a mechanism clearly different from void growth from vacancy condensation appearing under irradiation and known as swelling \cite{cawthorne,mansur}.} \textcolor{black}{Void growth and coalescence is still the predominant fracture mechanism of austenitic stainless steels irradiated in LWR conditions. Another mode of fracture - known as channel fracture - has also been reported for these steels under specific conditions (higher irradiation temperatures) but is not considered in this study \cite{margolin2016}.}
Regarding fusion applications, lower irradiation temperatures and doses are considered, leading to less pronounced decrease of fracture toughness of austenitic stainless steels \cite{alexander1995}. Materials selection of ITER reactor has required assessing in particular the fracture toughness of pure copper and copper alloys, irradiated and tested at relatively low temperatures (below 300 $^{\circ}$C). Copper alloys have been recently selected to be used in the final ITER blanket system. Significant hardening \textcolor{black}{(Fig.~\ref{fig0})} and reduction of uniform elongation for low doses are observed for pure copper and copper alloys\textcolor{black}{, as a result of the production of irradiation defects such as dislocation loops and Stacking Fault Tetrahedra (SFT)}. Fracture toughness of copper alloys has been shown to depend on temperature, and the decrease with irradiation strongly depends on alloying elements: significant decrease was observed for CuAl25 while no significant effect was observed for CuCrZr at low testing temperatures \cite{tahtinen,li}. Fractographic observations \cite{alexander99,li} indicate that room temperature fracture mechanisms involved microvoid coalescence.
A comprehensive review of mechanical properties of unirradiated and irradiated copper and copper alloys can be found in \cite{reviewcu}.

Empirical bounding curves of fracture toughness have been defined for engineering purposes (see \cite{chopra1} for irradiated austenitic stainless steels). Correlation of fracture toughness with tensile properties (evolution of yield stress and uniform elongation with irradiation) has also been proposed \cite{odette}, while reduction of uniform (or total) elongation measured on tensile tests does not correlate in general with reduction of fracture toughness. Physically-based models of ductile fracture through void growth and coalescence are now widely used for unirradiated materials, following seminal contributions of McClintock \cite{mcclintock} and Rice and Tracey \cite{ricetracey} describing the behavior of voids under mechanical loading, and Gurson \cite{gurson} and Thomason \cite{thomason90} homogenized models regarding void growth and coalescence, respectively. Recent reviews of these models can be found in \cite{benzergaleblond,pineaureview,BLNT}. Assuming that deformation mechanisms are similar (which hold true for low irradiation dose), such models can a priori be applied to irradiated materials, and decrease of fracture toughness $J_{Ic}$ with irradiation can be rationalized as resulting from loss of strain-hardening capability (decreasing $J_{Ic}$) and hardening (increasing $J_{Ic}$) \cite{pardoenhutchinson}, the former effect being dominant. Applications of these physically-based models to irradiated materials is more limited. Early models have been proposed in \cite{hamilton,odette} assuming microvoid coalescence. Physically-based models have been recently described (see \cite{margolin2014,margolin2016} and references therein) to describe fracture toughness of irradiated austenitic stainless steels, accounting for various phenomena such as void initiation, growth and coalescence, channelling, irradiation-induced nanovoids \cite{cawthorne}. More recently, void growth and coalescence has been assessed numerically \cite{lingJNM} at the crystal scale using physically-based constitutive equations developed for irradiated stainless steel \cite{hure2016}, showing accelerated growth and coalescence with irradiation.

Physically-based models aiming at predicting fracture toughness of irradiated materials \cite{margolin2014,margolin2016} assume some physical fracture mechanisms which need validation through dedicated experimental observations. \textcolor{black}{In particular, irradiation of austenitic stainless steels and copper alloys leads to a change at the crystal scale of deformation mechanisms from an homogeneous to heterogeneous one: dislocations sweep away irradiation defects in narrow channels, making subsequent motion of dislocations easier. Assessing the effects of these irradiation-hardening mechanisms on void growth and coalescence is therefore required.}
 The objective of this study is thus to assess experimentally void growth and coalescence in an irradiated material. Pure copper has been selected as a model FCC material to describe austenitic stainless steels for LWR applications (both sharing similar evolution of mechanical properties with irradiation). Moreover, void growth and coalescence in irradiated copper is relevant for fusion applications. Section~2 describes the material and methods used in this study. Analytical models are also presented. Section~3 details the experimental and numerical results, that are discussed in Section~4.

\section{Material and Methods}

\subsection{Material and Irradiation}

In addition to being relevant for fusion applications \cite{iter}, pure copper has been used in this study as a model FCC material. Compared to austenitic stainless steels more relevant for fission applications (LWR internals structure, FBR claddings), copper has also a Face-Centered Cubic crystallographic structure, shows a high sensibility to irradiation with significant hardening for low doses, and saturation of mechanical properties below 0.1dpa (compared to few dpa for SS), and high thermal conductivity \cite{reviewcu}. The two last properties simplify irradiation with ions, the former by reducing irradiation time, the latter by allowing to achieve high flux (and therefore reducing also irradiation time) with a good monitoring of temperature. 75~$\mu$m foils have been supplied by Goodfellow$ ^{\textregistered}$ with a typical chemical composition: \textcolor{black}{Cu $>99.9\%$, Ag 500 ppm, O 400 ppm, Bi $<$10 ppm, Pb $<$50 ppm, other metals $<$300 ppm}. Initially in an hardened state, foils have been heat-treated (200~$^{\circ}$C, 30~min, air cooling) to restore some ductility. The heat-treatment conditions are a compromise between ease of use (to manipulate tensile samples without damaging them) and ductility (to get homogeneous plastic strain (up to few percents) along the gauge length of tensile samples, \textit{i.e.} before necking). Electron Back Scattered Diffraction (EBSD) revealed that the material shows no texture, with a significant number of twins, and a mean grain size of about $20~\mu$m (Fig.~\ref{fig2}b).
\begin{figure}[H]
\centering
\subfigure[]{\includegraphics[height = 4.5cm]{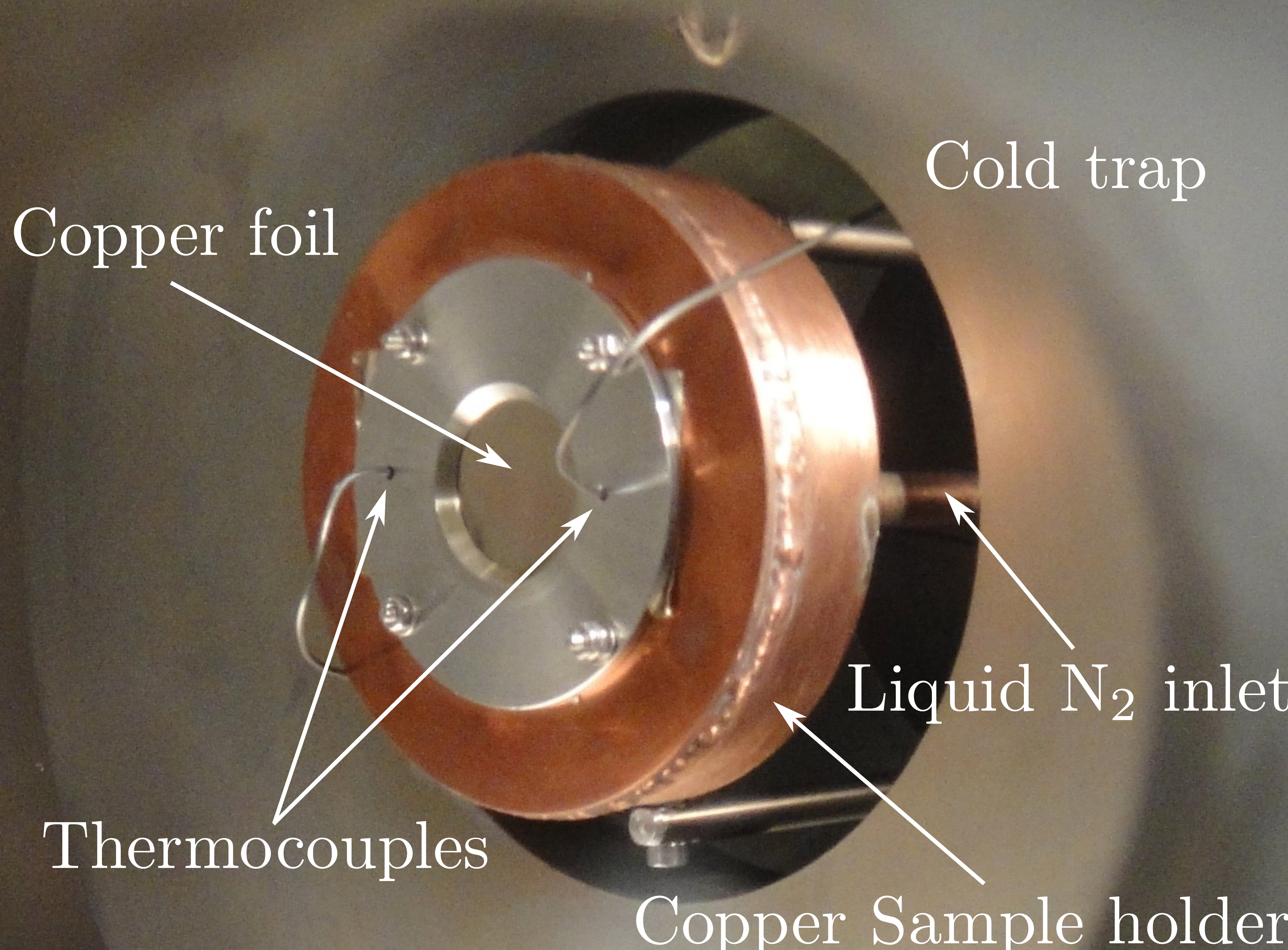}}
\hspace{0.5cm}
\subfigure[]{\includegraphics[height = 4.5cm]{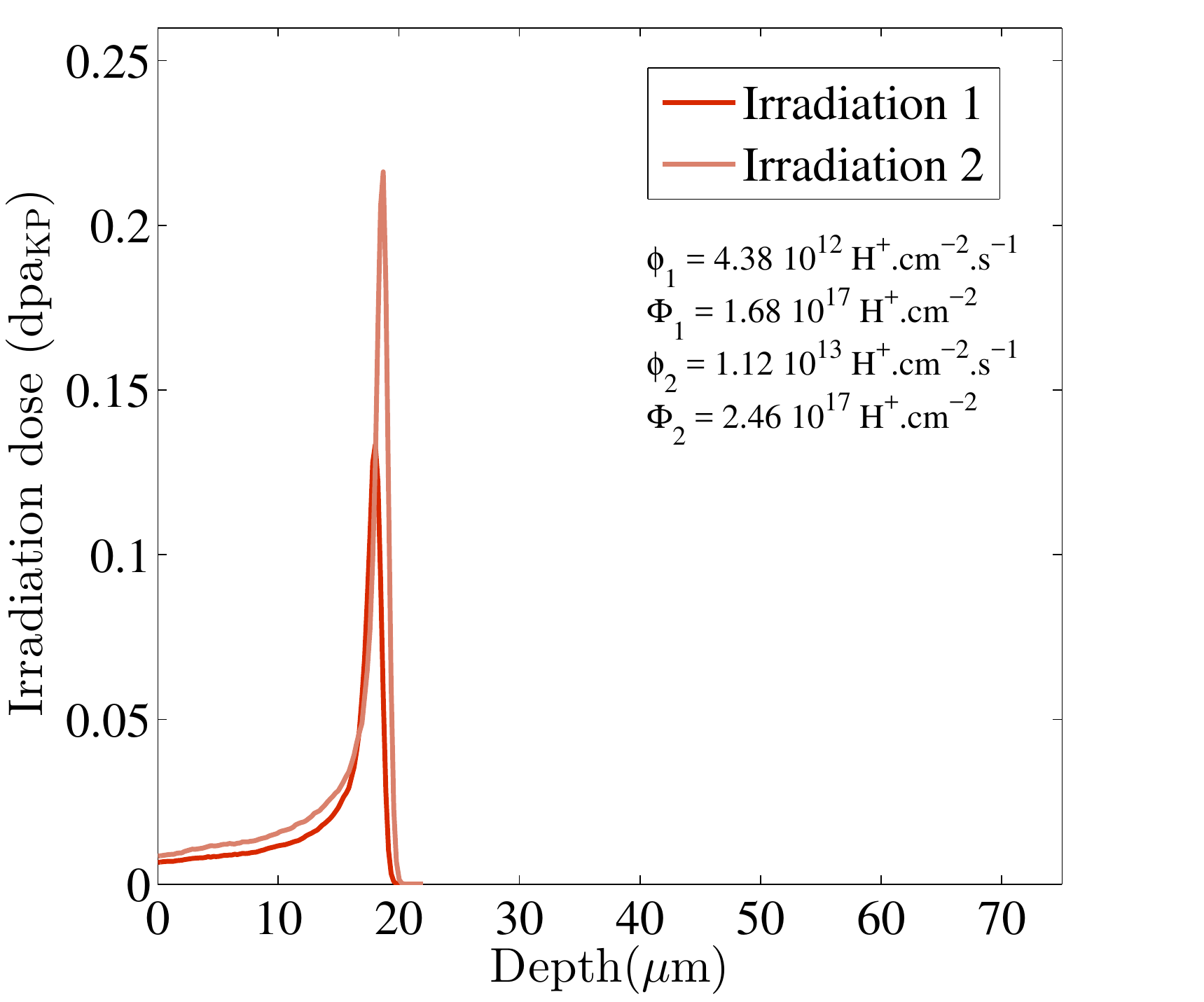}}
\subfigure[]{\includegraphics[height = 3.8cm]{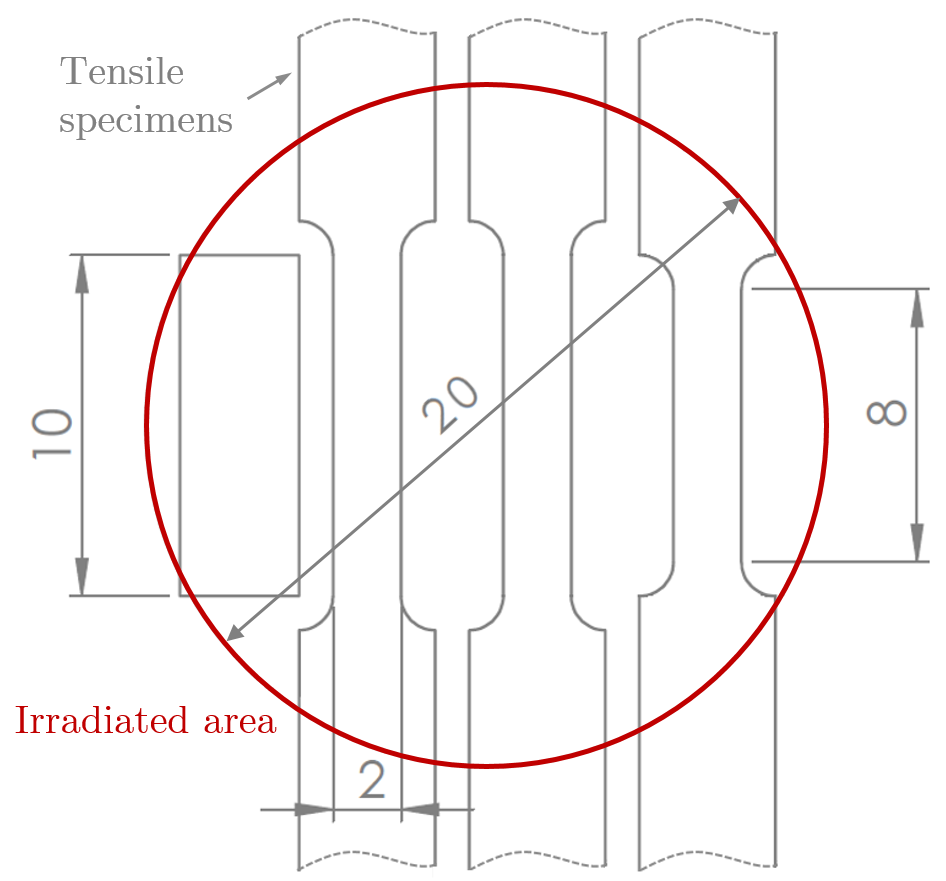}}
\caption{(a) \textcolor{black}{Irradiation setup: the copper foil is fixed to a copper sample holder cooled by liquid nitrogen} (b) Dose as a function of depth through the copper foils for the two irradiation performed. Dpa levels were computed using SRIM-2013 software, with Kinchin-Pease (KP) model, with a displacement energy of 30~eV. \textcolor{black}{(c) Sketch of the sampling of tensile specimens in the irradiated foil ensuring that the gauge length is located in the irradiated area circled in red (Dimensions in mm).}}
\label{fig1}
\end{figure}
Two copper foils were irradiated, a third one was kept to get reference data. Irradiations were performed at JANNuS facility (CEA, Saclay) \cite{jannus} with 2~MeV protons. Such energy ensures that a significant thickness of the material is irradiated  while avoiding the $\mathrm{^{65}Cu(p,n)^{65}Zn}$ nuclear reaction\footnote{The threshold energy of $\mathrm{^{65}Cu(p,n)^{65}Zn}$ is 2.17~MeV \cite{johnson,switkowski}.} making the samples radioactive. \textcolor{black}{Each copper foil was fixed to a copper sample holder cooled by liquid nitrogen (Fig.~\ref{fig1}a) in order to avoid any heating due to the energetic proton beam. A 20~mm diameter disk-shape region of the foils was irradiated, corresponding to the rastering of the millimetric ion beam \cite{jannus}}. Sample temperature was monitored to be below $20~\mathrm{^{\circ}C}$ throughout the irradiations. Flux and fluence obtained for each irradiation are given on Fig.~\ref{fig1}b. SRIM-2013 software \cite{SRIM} was used to compute Displacement Per Atom (thereafter noted \textit{dpa}), using Kinchin-Pease (KP) model \cite{stollerdpa} and a displacement energy of 30~eV for copper \cite{astmed}. For irradiation 1, ion-beam angle of $15^{\circ}$ with the foil normal was also accounted for in the calculations. Dpa levels as a function of depth are shown on Fig.~\ref{fig1}b where doses of about 0.015~dpa and 0.15~dpa are measured at the surface and at the Bragg peak, respectively. Irradiation depth is about $19~\mu$m, \textit{i.e.}, only one quarter of the foil thickness was irradiated. 

\subsection{Experimental set-up}

\begin{figure}[H]
\centering
\subfigure[]{\includegraphics[height = 5cm]{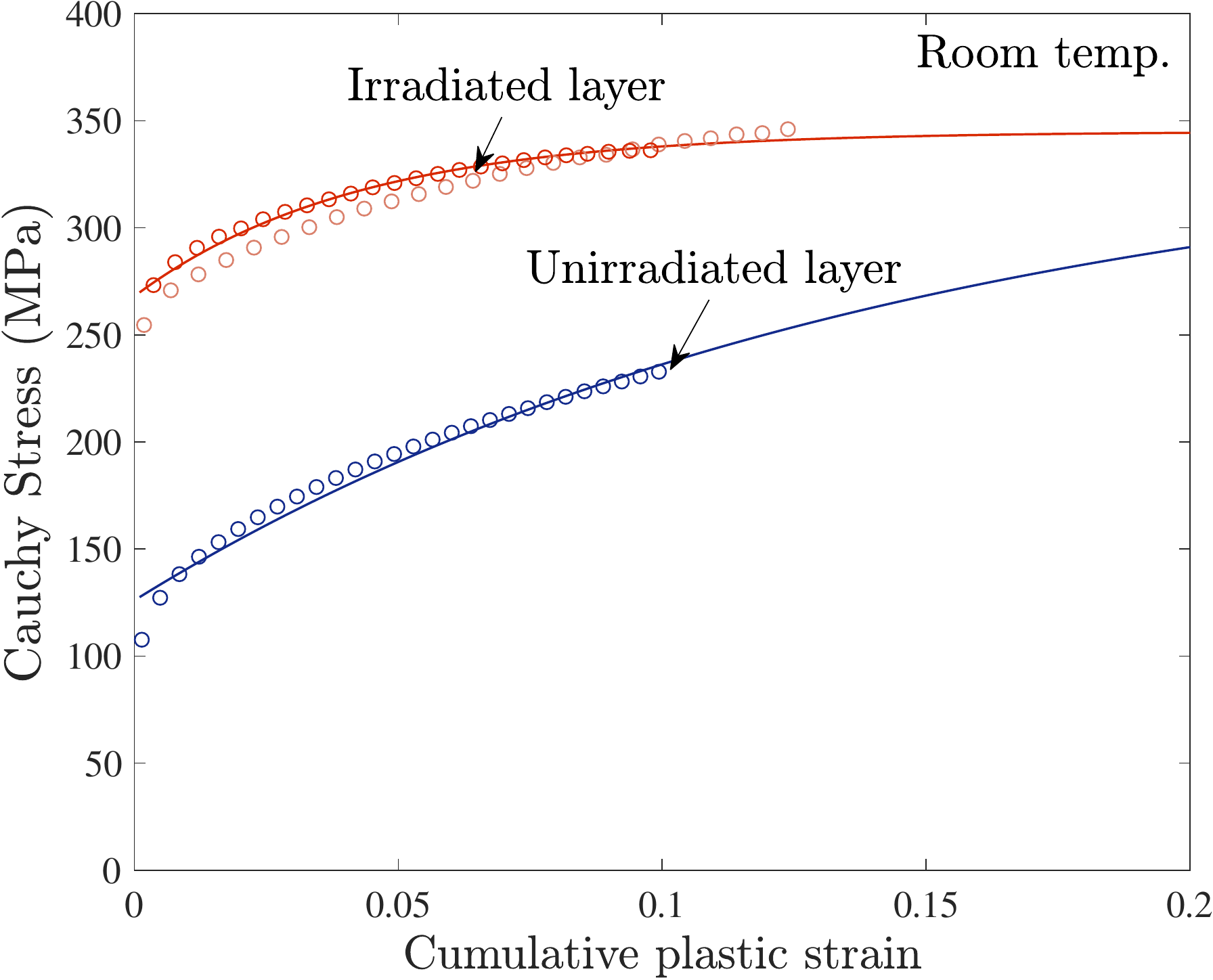}}
\hspace{1cm}
\subfigure[]{\includegraphics[height = 5cm]{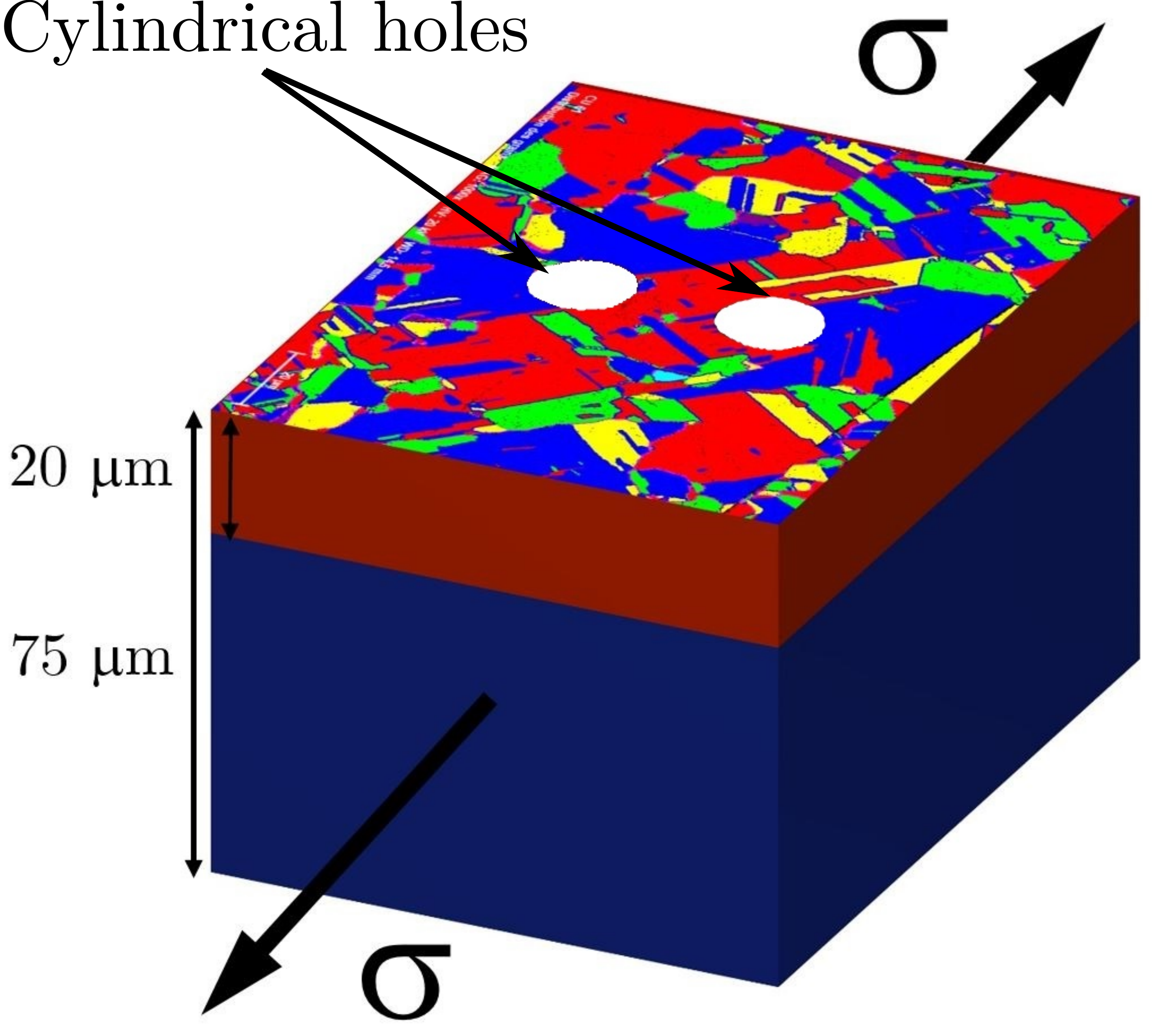}}
\caption{(a) Stress-strain curves of unirradiated and irradiated copper. Points correspond to the experimental data, lines to numerical results. (b) Sketch of the experimental setup: cylindrical voids are drilled through unirradiated and partly-irradiated copper foils and put under uniaxial tension.}
\label{fig2}
\end{figure}

\textcolor{black}{Tensile samples of gauge shape of 10mm in length and 2mm in width were machined from the unirradiated and irradiated foils by a conventional milling machine (Fig.~\ref{fig1}c)}. Tensile tests were performed \textcolor{black}{on a conventional electromechanical tensile machine, equipped with a 1~kN load cell,} at room temperature at a mean strain rate of $5.10^{-4}~\mathrm{s^{-1}}$. Conventional 0.2\% yield stress of the unirradiated material is equal to \textcolor{black}{110~MPa} (Fig.~\ref{fig2}a and Tab.~1). \textcolor{black}{Other conventional tensile properties such as tensile strength, uniform and total elongation are not reported as being dependent on the sample geometry, especially in the case of thin specimens.} Tensile tests of the irradiated foils do not lead directly to the tensile properties of the irradiated material, as the irradiated layer is only one quarter of the foil's thickness. However, the sole contribution of the irradiated layer on the measured force can be obtained, for a given applied strain, through subtracting the contribution of the unirradiated layer (known from the tensile tests on the unirradiated material). Assuming equal strains in the unirradiated and irradiated layers leads to the average stress-strain behavior of the irradiated layer plotted on Fig.~\ref{fig2}a. A significant hardening of the irradiated layer is observed (as detailed in Tab.~1), as well as a decrease of strain hardening capability. Both irradiations (Fig.~\ref{fig1}b) lead to very similar stress-strain behavior, so that they are are not differentiated in the following.\\

\begin{table}[H]
  \begin{tabular}{|c|c|c|c||c|c|}
    \hline
    Flux & Fluence & Dose & $T_{\mathrm{Irradiation}}$ & $T_{\mathrm{Test}}$ & $\mathrm{R_{p,0.2\%}}$ \\
    ($\mathrm{H^+.cm^{-2}.s^{-1}}$) & ($\mathrm{H^+.cm^{-2}}$) & (dpaKP) &  ($\mathrm{^{\circ}C}$) & ($\mathrm{^{\circ}C}$) & (MPa) \\ 
    \hline
    - & - & - & - & 20 & 110 \\
    $ 4.4\,10^{12} $  & $1.7\,10^{17}$   & [0.015 - 0.13]  & $<20$  & 20 & 254 \\
    $1.1\,10^{13} $  & $ 2.5\,10^{17}$   & [0.015 - 0.2]  & $<20$  & 20 & 260 \\
    \hline
  \end{tabular}
  \caption{\textcolor{black}{Conventional 0.2\% yield stress $\mathrm{R_{p,0.2\%}}$ of the unirradiated material and of the irradiated layer of the irradiated material. Dose range reported corresponds to the minimal and maximal values in the irradiated layer (Fig.~\ref{fig1}b).}}
  \label{tabyield}
\end{table}

The advantage of extracting the irradiated stress-strain behavior by testing a composite tensile specimen (irradiated / unirradiated) is to delay necking that would have happened at a lower strain if the tensile specimen was fully irradiated. This technique allows to assess the material behavior up to higher strain and/or for materials prone to localize, and was already used in the context of L\"uders bands \cite{hallai}. As shown on Fig.~\ref{fig0}, the yield stress of the irradiated layer found in this study is in quantitative agreement with values reported in the literature, albeit obtained for annealed copper and higher irradiation temperature, which indicates that this evolution may also be relevant for lower irradiation temperature and slightly cold-worked material. 

\begin{figure}[H]
\centering
\subfigure[]{\includegraphics[height = 4.5cm]{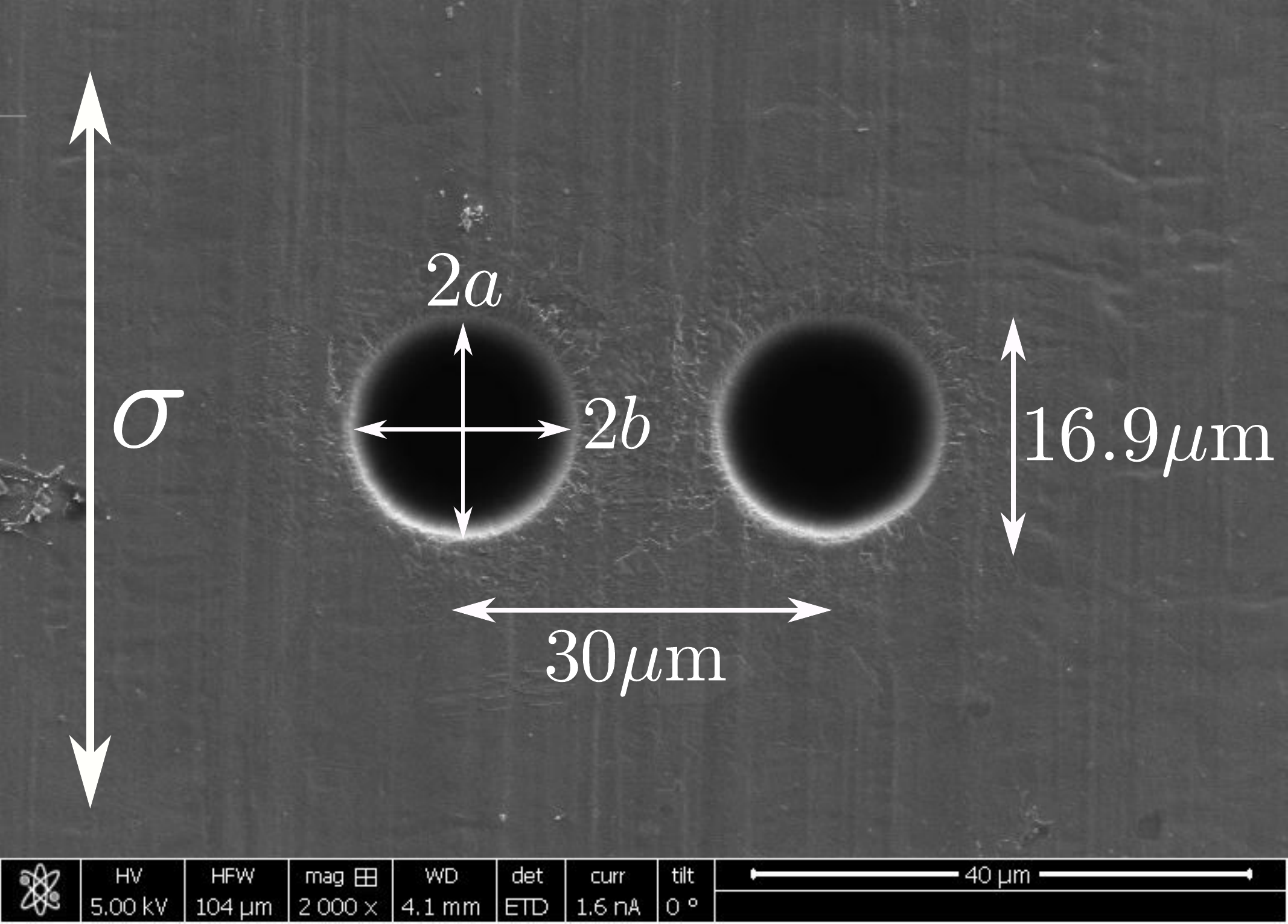}}
\subfigure[]{\includegraphics[height = 4.5cm]{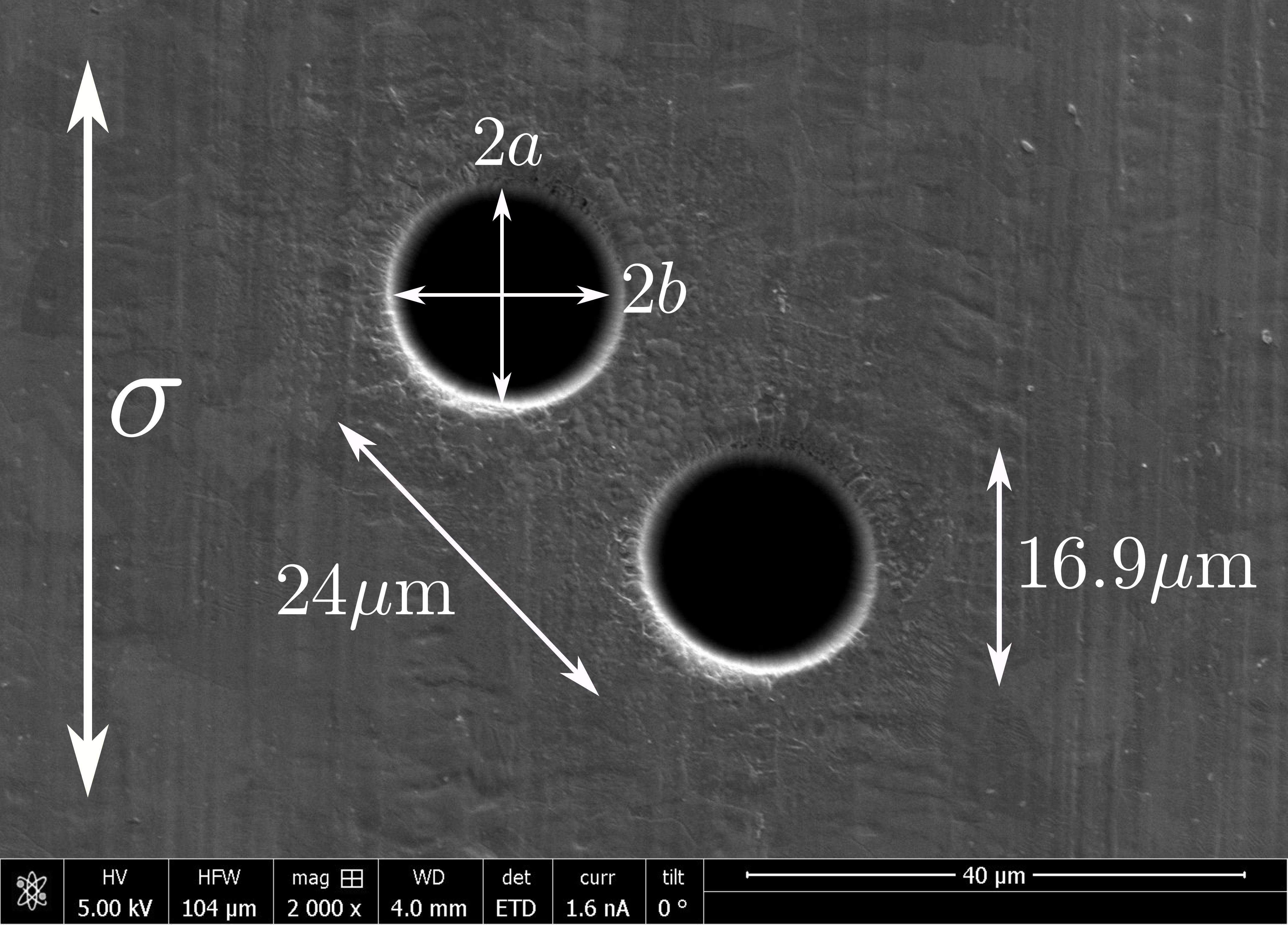}}
\caption{SEM observations of Focused-Ion Beam (FIB) drilled cylindrical voids through tensile samples. Two geometries are considered that differ from the orientation of the intervoid axis to the loading direction, referred to as 90$^{\circ}$ (a) and 45$^{\circ}$ (b).}
\label{fig3}
\end{figure}

In order to assess void growth and coalescence, model cylindrical voids were drilled using Focused-Ion Beam (FIB) atomic milling throughout the tensile samples (Fig.~\ref{fig2}b). Early experiments on void growth and coalescence using a similar methodology can be found in \cite{mcclintock}, while recent results have been obtained based on laser-drilling \cite{weck}. Advantages of FIB drilling include precise control of void geometry and absence of heat-affected zones requiring annealing. However, more time is needed for FIB drilling compared to laser-drilling. Two different configurations were selected, as shown on Fig.~\ref{fig3}, that differ from the orientation of the intervoid axis with respect to the loading direction. FIB milling leads to slightly conical through-thickness void shape that has been minimized by performing drilling on both sides on the specimens. Void diameter remains however slightly smaller in the middle of the specimen than on the surface. The mean diameter is $16.9~\mu$m and the intervoid distance is $30\mu$m for the configuration 90$^{\circ}$ and $24~\mu$m for the configuration 45$^{\circ}$ (Fig.~\ref{fig3}). \textcolor{black}{Void diameter was selected as the smallest allowing drilling completely through the thickness of the foils, while the inter-void distance was selected based on preliminary experiments in order to get void coalescence before tensile specimen failure.} Interrupted tensile tests were performed at room temperature up to a given value of macroscopic plastic strain where tensile specimens were unloaded and voids deformation observed under Scanning Electron Microscope (SEM). Void shape is described by measuring the semi-axis $a$ and $b$ of the ellipse inscribed in the deformed void. The procedure is repeated until coalescence occurs. For each set of parameters (voids geometry, unirradiated/irradiated materials), experiments were repeated at least 2 times, and average results are presented hereafter. 

\subsection{Analytical and numerical modeling}
\label{anum}
Experimental results on void growth and coalescence are compared in Section~\ref{resu} to both numerical simulations and analytical model. Finite elements simulations were performed using the finite element solver Cast3M \cite{castem} using finite strain elastoplastic constitutive equations implemented with the MFront code generator \cite{mfront}. 3D meshes of quadratic elements have been used, and boundary conditions ensuring uniaxial tension have been applied. For the irradiated specimen, both irradiated and unirradiated layers have been modeled with the different behaviors described below\textcolor{black}{\textit{, i.e.,} the composite structure of the irradiated specimen (Fig.~\ref{fig2}b) is fully accounted for by specifying different material parameters in the irradiated layer and in the unirradiated layer}. Mesh convergence was checked for all simulations presented hereafter. The voids size being of the order of the grain size, the material was modelled using classical time-independent plasticity, using von Mises yield criterion and isotropic hardening $R(p) = R_0 + Q_1[1-\exp(-b_1 p)]$, where $p$ is the cumulated plastic strain \textcolor{black}{(defined such as $\dot{p}^2 = [2/3]\underline{\dot{\varepsilon}}_p:\underline{\dot{\varepsilon}}_p$ with $\underline{\dot{\varepsilon}}_p$ the increment of plastic strain tensor)}, and $\{R_0,Q_1,b_1\}$ are material parameters. \textcolor{black}{$R_0$, $Q_1$ and $b_1$ are phenomenological parameters aiming at modeling the hardening of the material. More precisely, $R_0$ corresponds to the yield stress, $Q_1$ to the maximal hardening due to deformation, and $b_1$ to the hardening rate (with respect to strain)}. Hardening law was chosen in order to be able to reproduce key features such as initial yield point and saturation at high plastic strain, while keeping the number of parameters as low as possible. $(R_0 + Q_1)$ was fixed to $345$~MPa, corresponding to the saturation yield stress obtained for highly irradiated pure copper \cite{reviewcu}, in agreement with the stress saturation at high strain obtained through torsion tests. Parameters $R_0$ and $b_1$ were adjusted based on tensile tests for both unirradiated and irradiated materials, leading to $\{R_0=126~\mathrm{MPa},\  b_1=7\}$ and $\{R_0=268~\mathrm{MPa},\ b_1=24\}$, respectively. Elasticity is assumed to follow Hook's law, with Young's modulus $E=120$~GPa and Poisson's ratio $\nu=0.3$. With these parameters, a good agreement is observed between experimental tensile curves and numerical simulations, as shown on Fig.~\ref{fig2}a. 

The predictions of McClintock analytical cylindrical void growth model \cite{mcclintock}, \textcolor{black}{detailed in Appendix A}, were also compared to the experimental results. In the growth regime, \textit{i.e.}, when voids do not interact strongly with each other, each void is subjected to an uniaxial stress state, and McClintock model reads:
\begin{equation}
\left\{
\begin{aligned}
a + b    &= (a_0 + b_0)\left[1 + p \frac{1+2\sqrt{3}}{4}   \right] \\
a - b   &=  (a+b)\sqrt{3}p \\
\end{aligned}
\right.
\label{eq1}
\end{equation}
where the subscript $0$ corresponds to the initial values. In the coalescence regime, \textit{i.e.}, when voids interact strongly between each other, McClintock model is used assuming, at first approximation, that voids are subjected to equibiaxial stress state\footnote{Superposition of macroscopic uniaxial loading and interactions with adjacent voids lead to a more biaxial stress state during coalescence, which is approximated as an equibiaxial stress state.}:
\begin{equation}
\left\{
\begin{aligned}
a + b    &= (a_c + b_c)\exp{\left[(p -  p_c) (1 + \sqrt{3} \sinh{(\sqrt{3})}) \right]} \\
a - b   &=  (a+b)\frac{a_c - b_c}{a_c+b_c}\exp{\left[-2\sqrt{3} \sinh{(\sqrt{3})}(p -  p_c)     \right]} \\
\end{aligned}
\right.
\label{eq2}
\end{equation}
where $\{a_c,b_c\}$ and $p_c$ are the semi-axis of the voids and the plastic strain at the onset of coalescence, respectively. 

\section{Experimental and numerical results}
\label{resu}
\subsection{Typical SEM observations}

\begin{figure}[H]
\centering
\subfigure[]{\includegraphics[height = 6cm]{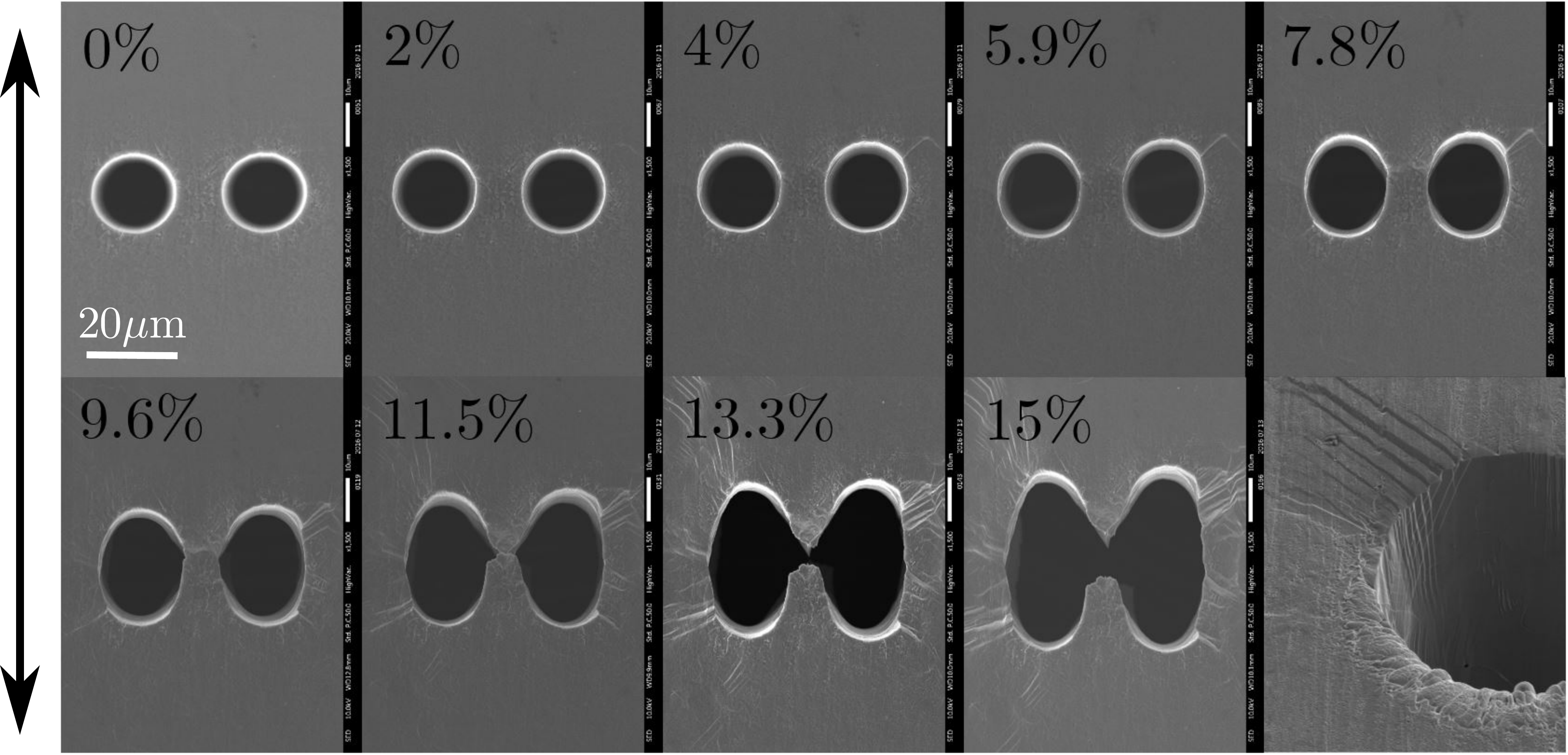}}
\subfigure[]{\includegraphics[height = 6cm]{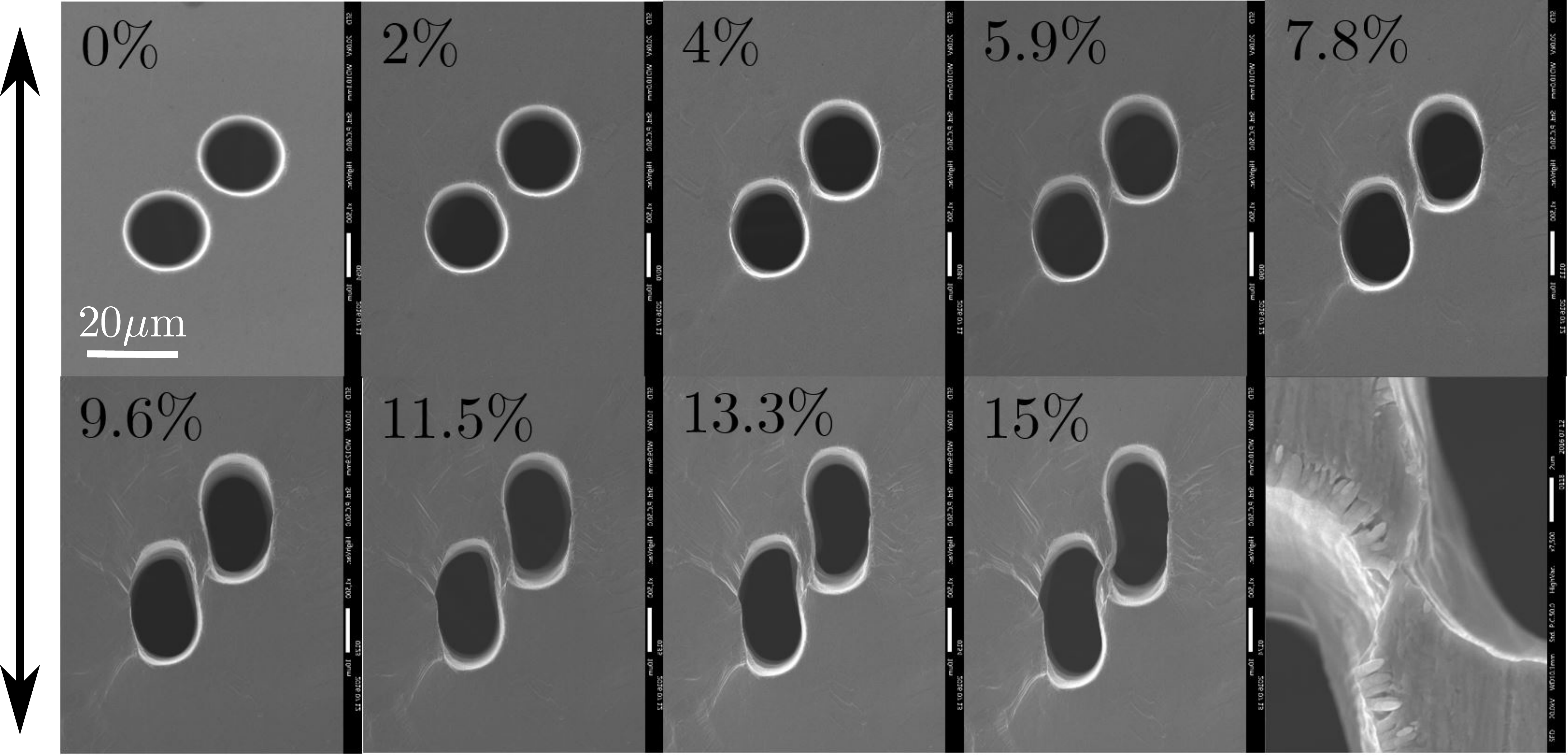}}
\caption{Typical SEM observations of the evolution of \textcolor{black}{void (hole)} shapes as a function of applied plastic strain at the irradiated state. \textcolor{black}{Tensile direction is indicated by the arrow.} Last picture shows details of (a) slip bands at the void edge and (b) shear band leading to void coalescence \textcolor{black}{at the irradiated state}.}
\label{fig4}
\end{figure}

Typical SEM observations of void shapes as a function of applied plastic strain are shown on Fig.~\ref{fig4}. Voids start to elongate in the loading direction, associated with a contraction in the perpendicular direction. After a critical strain (that will be referred and defined in Section~\ref{resunum} to as onset of coalescence), void elongation perpendicular to the loading direction starts, ultimately leading to void coalescence. Slip bands are observed in the highly deformed regions close to the voids, especially for plastic strain higher than 10$\%$. Coalescence appears mainly through internal necking for the $90^{\circ}$ configuration, while localization through shear band occurs in the $45^{\circ}$ configuration

Void growth and coalescence were found to be qualitatively similar on unirradiated samples and on the irradiated sides of irradiated samples\footnote{\textcolor{black}{Due to the composite structure of the irradiated samples (Fig.~\ref{fig2}a), hole shapes are different on both sides of the samples. More precisely, the hole shape of the unirradiated side of the irradiated sample is the same as the hole shape of the unirradiated sample, for a given plastic strain.}}. However, some differences were observed regarding grain scale plasticity which appears more heterogeneous after irradiation as shown from marked slips bands at the surface (\textit{e.g.}, Fig.~\ref{fig4}a). The potential effect of heterogeneous plasticity induced by irradiation on void growth and coalescence is assessed in Section~\ref{resunum} where numerical simulations that do not account for heterogeneous plasticity are compared to experimental data.

\subsection{Experimental results \textit{vs.} numerical modeling}
\label{resunum}

\begin{figure}[H]
\centering
\subfigure[]{\includegraphics[height = 5cm]{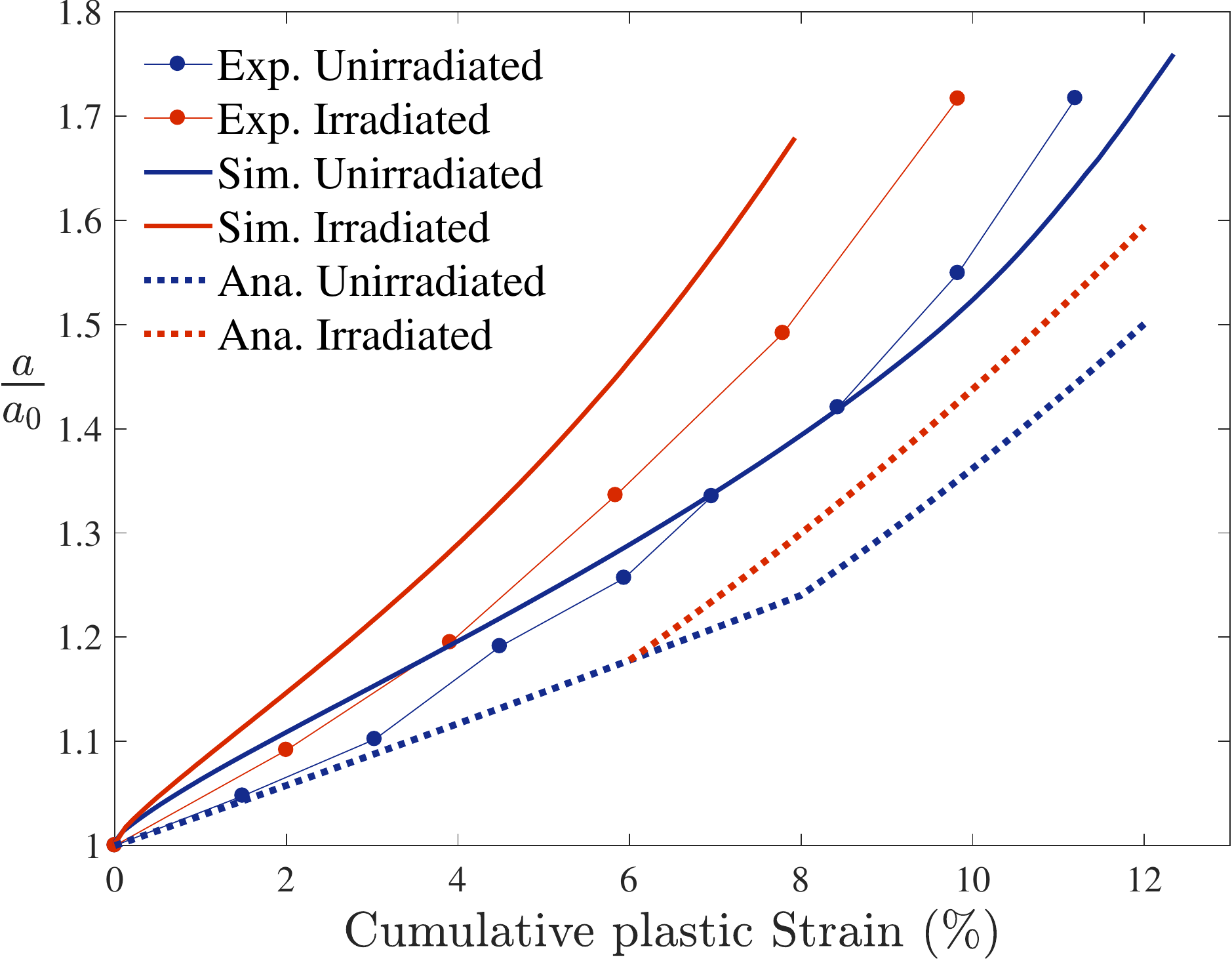}}
\hspace{0.5cm}
\subfigure[]{\includegraphics[height = 5cm]{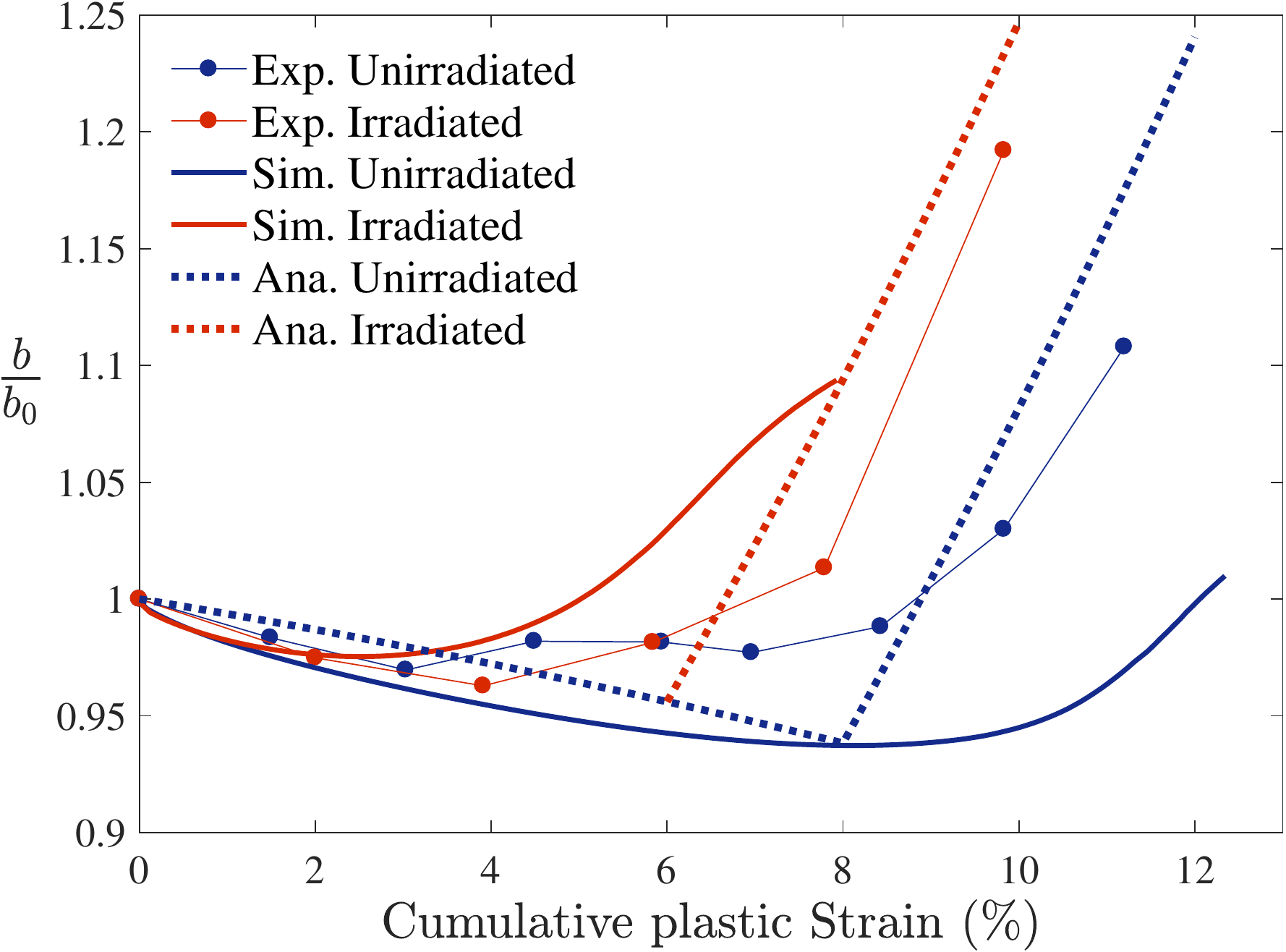}}
\caption{Evolution of voids axis $a$ and $b$ \textcolor{black}{(Fig.~\ref{fig3})} \textcolor{black}{normalized by their initial value $a_0$ and $b_0$ (with $a_0 = b_0 = 8.5~\mu$m)} as a function of applied plastic strain for the 90$^{\circ}$ configuration. Comparisons of experimental data (points) to numerical simulations (solid lines) and analytical predictions (dashed lines).}
\label{fig5}
\end{figure}

\begin{figure}[H]
\centering
\subfigure[]{\includegraphics[height = 5cm]{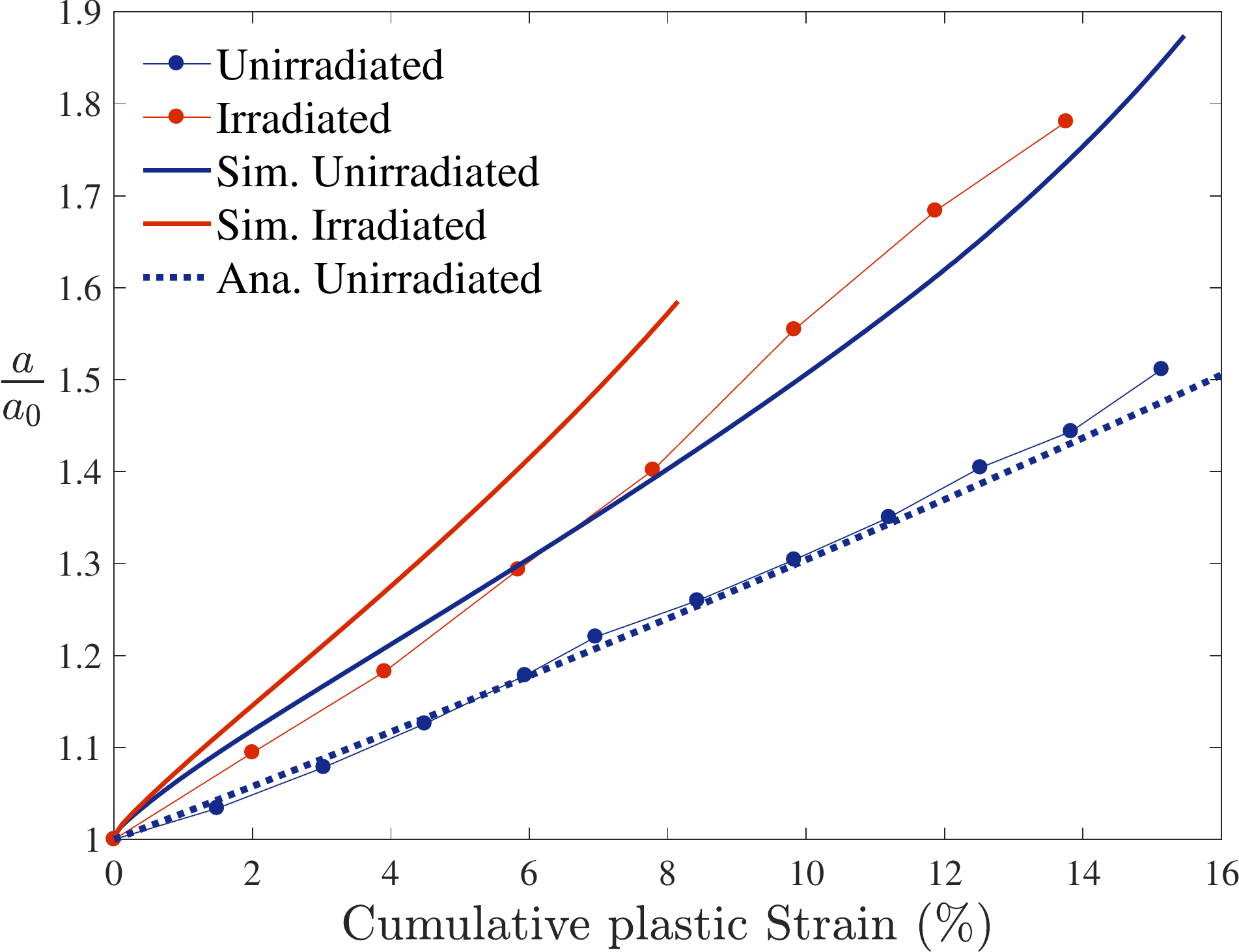}}
\hspace{0.5cm}
\subfigure[]{\includegraphics[height = 5cm]{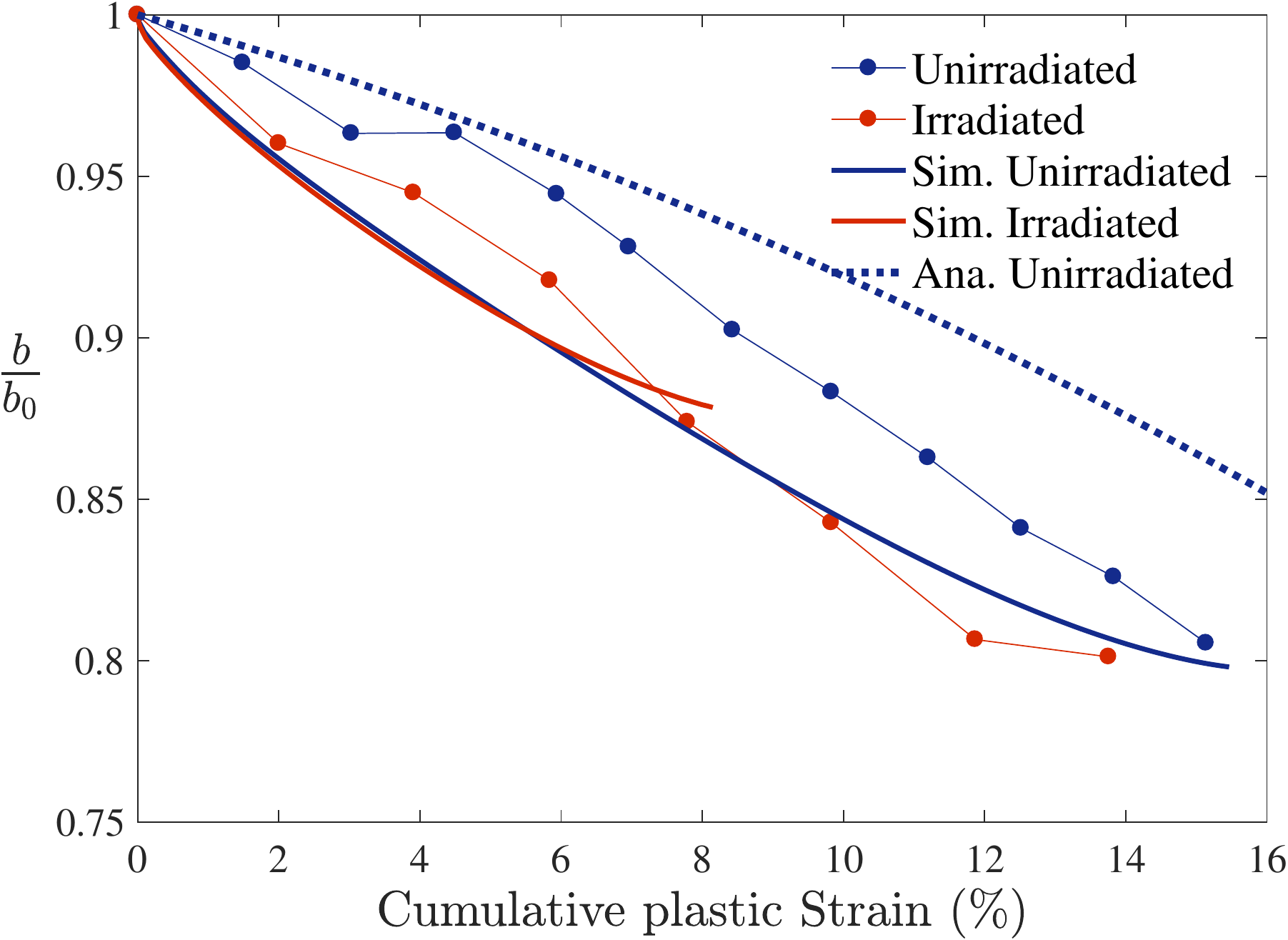}}
\caption{Evolution of voids axis $a$ and $b$ \textcolor{black}{(Fig.~\ref{fig3})} \textcolor{black}{normalized by their initial value $a_0$ and $b_0$ (with $a_0 = b_0 = 8.5~\mu$m)} as a function of applied plastic strain for the 45$^{\circ}$ configuration. Comparisons of experimental data (points) to numerical simulations (solid lines) and analytical predictions (dashed lines).}
\label{fig6}
\end{figure}

\begin{figure}[H]
\centering
\subfigure[]{\includegraphics[height = 5cm]{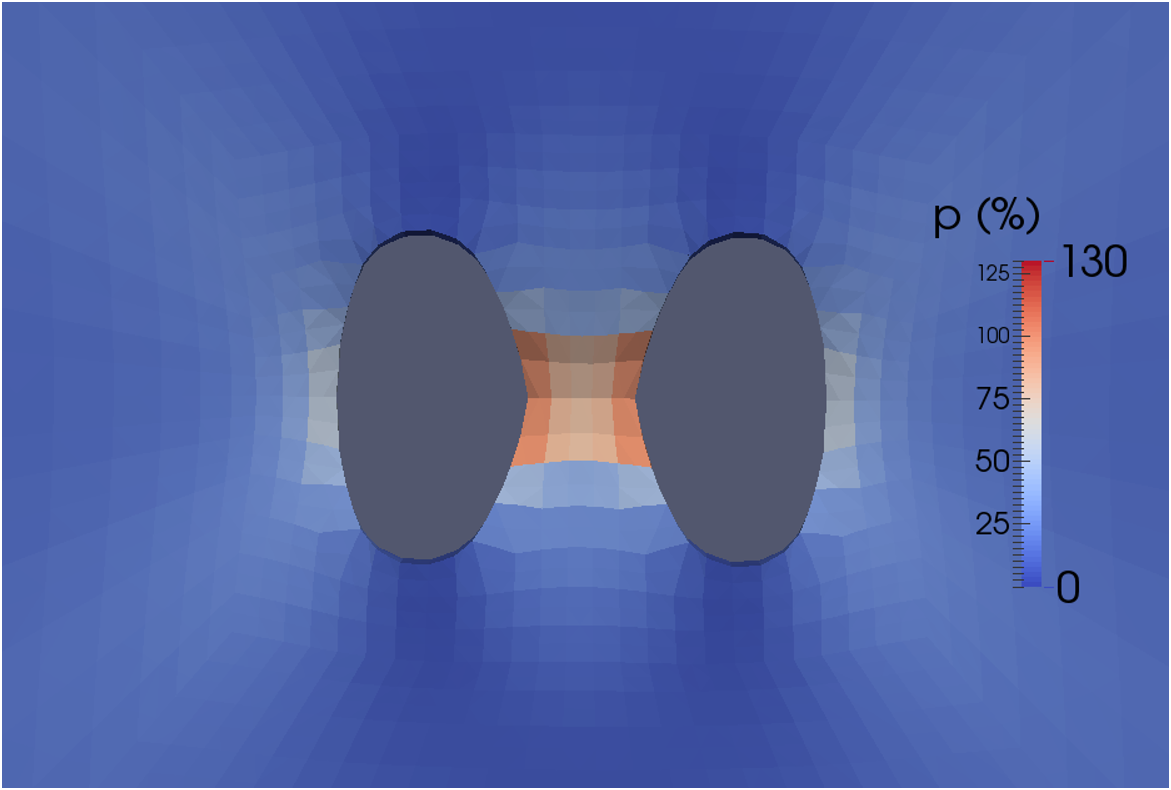}}
\hspace{0.5cm}
\subfigure[]{\includegraphics[height = 5cm]{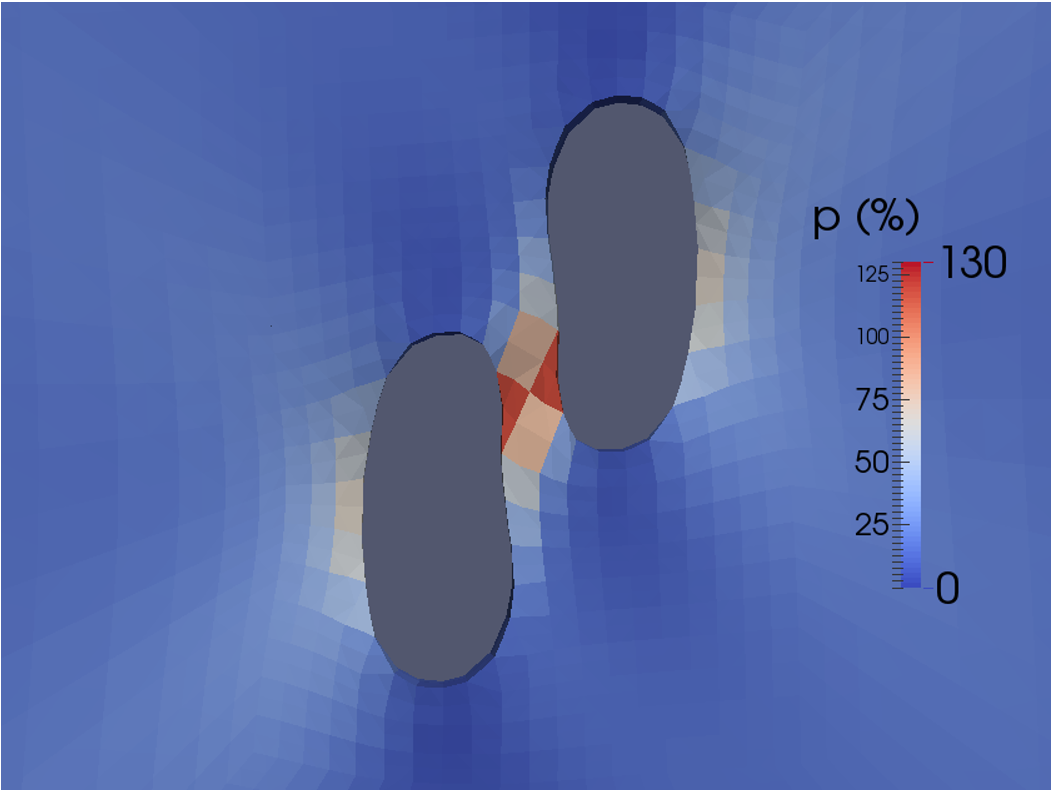}}
\caption{Numerical simulations of void growth and coalescence: local plastic strain field \textcolor{black}{p as defined in Section~2.3} for the 90$^{\circ}$ (a) and 45$^{\circ}$ configurations.}
\label{fig7}
\end{figure}

The evolutions of voids axis $a$ and $b$ \textcolor{black}{(Fig.~\ref{fig3})} \textcolor{black}{normalized by their initial value $a_0$ and $b_0$ (with $a_0 = b_0 = 8.5~\mu$m)} as a function of applied plastic strain  are plotted on Fig.~\ref{fig5} and Fig.~\ref{fig6}. Each experimental data point corresponds to an average value computed on multiple experimental realizations. Experimental results were found to be sensitive to the geometry of the voids, especially to the intervoid distance. Therefore, only similar configurations (voids radius and intervoid distance) were selected for averaging. In the 90$^{\circ}$ configuration (Fig.~\ref{fig5}), voids elongate along the tensile direction and contract along the perpendicular direction, for low applied strain. Voids in the irradiated specimens were found to deform slightly more than for unirradiated specimens. After a critical strain - that will be referred to as the onset of coalescence $p_c$ - voids start to expand perpendicular to the loading direction (Fig.~\ref{fig5}b). This corresponds to the transition between the growth phase - where voids deform without interacting with each others - to the coalescence phase - where strong interactions between voids lead to localized plastic flow in the intervoid ligament \cite{koplik}. This coalescence mode is called internal necking. Voids were found to coalesce earlier in irradiated material than in unirradiated material: while the onset of coalescence strain is about $8\%$ at the unirradiated state, it decreases to about $4\%$ at the irradiated state. As a consequence, final failure occured earlier for voids in the irradiated material. 
For the 45$^{\circ}$ configuration (Fig.~\ref{fig6}), voids elongate in the tensile direction and contract perpendicularly up to failure, for both unirradiated and irradiated material. Contrary to the 90$^{\circ}$ configuration where voids coalesce by internal necking, coalescence (and void linkage) appears due to a highly localized deformation band as can be seen on Fig.~\ref{fig4}b. However, similar conclusion can be drawn that voids deform and coalesce faster in the irradiated material. Finite-element simulations, \textcolor{black}{that corresponds to solid lines in Figs.~\ref{fig5} and \ref{fig6}}, are found to be in good agreement with experimental data for both configurations, reproducing the faster deformation as a function of applied strain and earlier coalescence for irradiated material. The agreement is found to be satisfactory for 90$^{\circ}$ configuration without any fitting parameters\footnote{as the material parameters were determined from literature data and adjustment of tensile curves, see Section~\ref{anum}.}, while being only qualitative for the 45$^{\circ}$ configuration. Snapshots of numerical simulations are shown on Fig.~\ref{fig7}, where a good agreement is observed regarding void shapes and coalescence mode (internal necking \textit{vs.} localization). McClintock analytical model (Eqs.~\ref{eq1},~\ref{eq2}), \textcolor{black}{that correspond to dashed lines in Figs.~\ref{fig5} and \ref{fig6},} is found to reproduce the trends observed in experimental results. 

Differences between experimental results and simulations can be attributed to different factors. First, the experimental methodology, while allowing to assess easily the physical mechanisms involved, has been found to be rather sensitive to the geometry considered (void radius, distance between voids). Simulations have been performed to be as close as possible to the experiments, but the fact that FIB drilling leads to slightly conical voids (through the thickness) may have an effect. More importantly, void coalescence is sensitive to the strain-hardening behavior and is triggered once the strain-hardening modulus falls below a given value. Hence, adjusting the parameters of the hardening law can lead to an almost perfect agreement with experimental results. Here a choice has been made to adjust separately the parameters based only on tensile curves and informations about stress saturation at high strain, which leads clearly to get the trends but not a fully quantitative agreement.

\section{Discussion}

\textcolor{black}{Void growth and coalescence fracture mechanism has been widely assessed experimentally on both real materials - where voids come from inclusions cracking or decohesion - and more recently on model materials - where voids are precisely created on purpose - for unirradiated material (see \cite{pineaureview} for a review). By permitting a precise control of the geometry of the voids, model experiments allow assessing quantitatively the effect of the hardening behavior of the material around the voids on their subsequent growth and coalescence under deformation. Recent studies have for example described in details the effect of a pre-strain and work-hardening \cite{alinaghian} or the effect of local microstructure \cite{nemcko} on void growth and coalescence in unirradiated material. Similarly, irradiation-induced hardening is expected to have an effect that requires to be quantified. In this study,}
 experiments indicated that void growth and coalescence is accelerated in the irradiated material, which is qualitatively consistent with the decrease of fracture toughness reported in the literature as well as with what is observed in numerical simulations \cite{lingJNM}. These observations are in fact compatible with a coalescence criterion based on a critical void size $(a/a_0)_c$ ($\approx$ 1.4 for data presented in Fig.~5) valid for both unirradiated and irradiated material. Such kind of criterion has already been used for unirradiated materials \cite{besson2010} and can be used for irradiated material as a first approximation. Fractographic observations of (un-)irradiated stainless steels often reveal the presence of dimples smaller that the grain size, indicating that void growth and coalescence happen in fact in single crystals. The extreme case of this situation corresponds to highly-swollen materials where nano-voids contribute to fracture \cite{neustroev}. Void growth and coalescence has been studied at the crystal scale, leading to homogenized models (see, \textit{e.g.}, \cite{xuhan,ling,mbiakop}). Application of these models to irradiated materials requires to check experimentally if, at this scale, dislocation channelling - not included in these models - should be accounted for void growth and coalescence. In particular, intragranular voids are expected to have size on the order or below the typical distance between channels. Experiments on materials irradiated at high doses for which dislocation channelling is expected to be more pronounced are underway and will be presented elsewhere.\\

\section{Conclusions}

Irradiation has been shown to lead to significant modifications of mechanical properties of metals alloys. In particular, a decrease of fracture toughness of austenitic stainless steels used in LWR is observed \cite{epri}. Fracture mechanisms are usually inferred from fractographic observations, \textcolor{black}{and void growth and coalescence has been shown to be still predominant after irradiation for LWR's internals structures.}
A methodology used for unirradiated materials to assess void growth and coalescence mechanisms was adapted in this study \textcolor{black}{for the first time on ion-irradiated materials} and used on pure copper, taken as a model FCC material but also relevant for fusion applications \cite{reviewcu}. \textcolor{black}{The dose level used for copper ($\sim 0.01$~dpa) leads to a hardening comparable to the one observed for austenitic stainless steels at few dpa.} SEM observations showed similar growth and coalescence mechanisms for unirradiated and 0.015~dpa irradiated copper \textcolor{black}{but with an accelerated growth in the irradiated material, consistent with the decrease of fracture toughness reported}.  Numerical simulations have been performed considering only the hardening and decrease of strain hardening capability for the irradiated material. A good qualitative agreement was found between numerical simulations and experimental data. Thus, for the irradiation dose and micron-scale void size considered in this study, dislocation channeling, \textit{i.e.} heterogeneous deformation mode at the grain scale, does not play at first order a role in fracture mechanisms. This implies that classical ductile fracture models - that assume homogeneous deformation around voids - can be used for low to medium irradiation dose, justifying the use of homogenized ductile fracture models presented in \cite{benzergaleblond}.\\

\noindent
\textbf{Acknowledgments}\\

The authors would like to thank JANNuS Saclay team for performing the irradiations, Damien Schildknecht for machining the tensile samples and Fran\c coise Barcelo for the EBSD measurements. This project has received funding from the Euratom research and training programme 2014-2018 under Grant Agreement N$^{\circ}$661913. This work reflects only the authors' view and the Commission is not responsible for any use that may be made of the information it contains.

\section{Appendix A}
The original McClintock model predicting the deformation of cylindrical holes (of semimajor and semiminor axes $a$ et $b$) in a plastic material is \cite{mcclintock}:
\begin{equation}
\begin{aligned}
\ln{\frac{a + b}{a_0 + b_0}} &= \frac{p\sqrt{3}}{2(1-n)} \sinh{\left( \frac{\sqrt{3}(1-n)}{2} \frac{\sigma_a + \sigma_b}{\sigma_0}\right)} + \frac{\epsilon_a + \epsilon_b}{2} \\
m = \frac{a - b}{a + b} &= \frac{\sigma_a - \sigma_b}{\sigma_a + \sigma_b} + \left(m_0 - \frac{\sigma_a - \sigma_b}{\sigma_a + \sigma_b}     \right)\exp{\left[-\frac{\sqrt{3}p}{1-n} \sinh{\frac{\sqrt{3}(1-n)}{2} \frac{\sigma_a + \sigma_b}{\sigma_0}}      \right]}
\end{aligned}
\label{eqmc}
\end{equation}
where subscript $0$ refers to the initial configuration, $n$ is the strain-hardening coefficient of the material, and subscripts $a$ and $b$ refer to the direction of semimajor and semiminor axes. $\sigma$ and $\epsilon$ are the far field stress and strain, respectively. The parameter $n$ has only a weak effect on the predictions of McClintock model, thus is taken as $n=0$ in the following (which corresponds to perfectly plastic material).\\

\noindent
Under uniaxial tension ($\sigma_a = \sigma_0$, $\sigma_b = 0$, $\epsilon_a = p$, $\epsilon_b = -p/2$ ) of an initially circular hole ($a_0 = b_0$), Eq.~\ref{eqmc} becomes:
\begin{equation}
\begin{aligned}
\ln{\frac{a + b}{a_0 + b_0}} &= p\left(\frac{1 + 2\sqrt{3}\sinh{[\sqrt{3}/2}]}{4}   \right) \\
\frac{a - b}{a + b} &= 1 - \exp{\left(-\sqrt{3}\sinh{[\sqrt{3}/2]}p \right)}
\end{aligned}
\end{equation}
which reduces to Eq.~\ref{eq1} (taking $\sinh{[\sqrt{3}/2]} \approx 1$) and considering only the first term of the Taylor expansion (as plastic strain is rather low in the experiments).\\

\noindent
Under equibiaxial tension ($\sigma_a = \sigma_b = \sigma_0$, $\epsilon_a = \epsilon_b = p$ ) of an initially elliptical hole ($a_0 \neq b_0$), Eq.~\ref{eqmc} becomes:
\begin{equation}
\begin{aligned}
\ln{\frac{a + b}{a_0 + b_0}} &= p\left(1 + \sqrt{3}\sinh{[\sqrt{3}]}   \right) \\
\frac{a - b}{a + b} &= \frac{a_0 - b_0}{a_0 + b_0} \exp{\left(-2\sqrt{3}\sinh{[\sqrt{3}]}p \right)}
\end{aligned}
\end{equation}
which reduces to Eq.~\ref{eq2} considering also the first term of the Taylor expansion.\\
\bibliographystyle{elsarticle-num.bst}
\bibliography{spebib2}

\end{document}